# MCSAE: Masked Cross Self-Attentive Encoding for Speaker Embedding

*Soonshin Seo, Ji-Hwan Kim\**

Dept. of Computer Science and Engineering, Sogang University, Seoul, South Korea
{ssseo,kimjihwan}@sogang.ac.kr

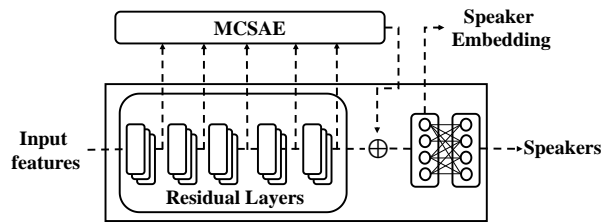

Figure 1: *Overview of proposed network using MCSAE*

## Abstract

In general, a self-attention mechanism has been applied for speaker embedding encoding. Previous studies focused on training the self-attention in a high-level layer, such as the last pooling layer. However, the effect of low-level features was reduced in the speaker embedding encoding. Therefore, we propose masked cross self-attentive encoding (MCSAE) using ResNet. It focuses on the features of both high-level and low-level layers. Based on multi-layer aggregation, the output features of each residual layer are used for the MCSAE. In the MCSAE, cross self-attention module is trained the interdependence of each input features. A random masking regularization module also applied to preventing overfitting problem. As such, the MCSAE enhances the weight of frames representing the speaker information. Then, the output features are concatenated and encoded to the speaker embedding. Therefore, a more informative speaker embedding is encoded by using the MCSAE. The experimental results showed an equal error rate of 2.63% and a minimum detection cost function of 0.1453 using the VoxCeleb1 evaluation dataset. These were improved performances compared with the previous self-attentive encoding and state-of-the-art encoding methods.

## 1. Introduction

Speaker recognition aims to identify speaker information from input speech. A type of speaker recognition is speaker verification (SV). It determines whether the test speaker's speech is accept or reject compared to the enrolled speaker's speech.

Traditionally, the Gaussian mixture model (GMM) with universal background model was proposed to encode supervector representing speaker information [1, 2]. Next, a joint factor analysis method was proposed to separate the supervector from the channel and speaker factors [3]. However, these methods are required an enormous amount of data for the enrollment. A disadvantage of the GMM-based supervector is its rapidly increasing dimensions depending on the Gaussian mixture. To solve this issue, an i-vector encoding method was proposed. It is capable of enrollment using one utterance, along with performing probabilistic linear discriminant analysis [4, 5].

Since the introduction of deep learning, d-vector was extracted directly from deep neural networks (DNNs) [6]. It is trained by using the DNN-based speaker classifier. Then the activations of the last hidden layer are encoded as speaker embedding. In addition, speaker embedding encoding methods using various DNN-based models were proposed. In time-delayed neural network (TDNN)-based model, the x-vector was proposed. It is a fixed dimensional statistics vector, encoded by using statistical pooling [7]. Among the convolutional neural network (CNN)-based models, ResNet [8] was used as a representative model for speaker embedding [9–14].

Attention mechanisms successfully applied to other areas, such as image and language processing [15–19]. In SV, TDNN or CNN model-based speaker embedding encoding methods using attention mechanism were proposed [9, 12, 20–26]. Especially, the self-attention mechanism [16] exhibited high performance in speaker embedding encoding method as called self-attentive pooling (SAP) [9, 24, 25]. The SAP is used to encode frame-level features into a utterance-level feature. It focuses on the frames by training interdependence with a context vector. In addition, an SAP-derived method called multi-head attentive pooling (MHAP) was proposed to improve performance [25].

However, these previous methods are focused on training the self-attention in a high-level layer instead of the lower-level layers. In other words, speaker embedding is encoded by using only the output feature of the last pooling layer. It results in decreased low-level features effect in the encoding of a speaker embedding. Therefore, it is difficult to encode the speaker embedding with an enhanced discriminative power.

Therefore, we propose a masked cross self-attentive encoding (MCSAE). This is a new SAP-derived speaker embedding encoding method using ResNet. MCSAE focuses on the features of both the high-level and low-level layers in the self-attention training. Based on multi-layer aggregation (MLA) [14], the output features of each residual layer are used as the input pair of the MCSAE, as shown in Figure 1. In the MCSAE, cross self-attention module is trained the interdependence of each input features. A random masking regularization module also applied to preventing overfitting problem. As such, the MCSAE enhances the weight of frames representing the speaker information. Then, the output features are concatenated and encoded to the speaker embedding. Therefore, a more informative speaker embedding is encoded by using the MCSAE.

We introduce the concept of self-attention and its use in Section 2, describe the proposed MCSAE method in Section 3, present the results in Section 4, and present our conclusions in Section 5.

---

\* Corresponding author

## 2. Concept of Self-Attention and Its Use

### 2.1. Self-attention mechanism

The principle of the self-attention mechanism is to focus on training the specific context information. In machine translation, self-attention using scaled dot-product attention and MHAP were proposed [16]. The formula for the scaled dot-product attention is

$$attention(\boldsymbol{Q}, \boldsymbol{K}, \boldsymbol{V}) = softmax\left(\frac{\boldsymbol{Q}\boldsymbol{K}^T}{\sqrt{d_k}}\right)\boldsymbol{V}. \quad (1)$$

The inputs are comprised of the query vector ($\boldsymbol{Q}$), key vector ($\boldsymbol{K}$), and value vector ($\boldsymbol{V}$). To train the relationship between $\boldsymbol{Q}$ and $\boldsymbol{K}$, scaling is applied to compute similarity using dot product operations on all $\boldsymbol{Q}$ and $\boldsymbol{K}$ elements and each element is divided by $\sqrt{d_k}$ ($d_k$ is the dimension of $\boldsymbol{K}$). Next, after applying the softmax method for normalization, the weights for $\boldsymbol{V}$ are obtained. The more similar $\boldsymbol{V}$ is to $\boldsymbol{Q}$, the higher its value, and thus, more attention will be paid to $\boldsymbol{V}$.

### 2.2. Self-attention in speaker verification

In SV, SAP, which is applied to TDNN and ResNet-based models, outperforms both the conventional temporal average pooling (TAP) and global average pooling (GAP) [9, 24, 25].

An input feature of the hidden layer $\boldsymbol{X} = [\boldsymbol{x}_1, \boldsymbol{x}_2, \ldots, \boldsymbol{x}_l, \ldots, \boldsymbol{x}_L]$ of length $L$ is fed into a fully-connected hidden layer to obtain $\boldsymbol{H} = [\boldsymbol{h}_1, \boldsymbol{h}_2, \ldots, \boldsymbol{h}_l, \ldots, \boldsymbol{h}_L]$. Given that $\boldsymbol{h}_l \in \mathbb{R}^d$ and a learnable context vector $\boldsymbol{u} \in \mathbb{R}^d$ the attention weight $w_l$ is measured by training the similarity between $\boldsymbol{h}_l$ and $\boldsymbol{u}$ with softmax normalization as

$$w_l = \frac{exp(\boldsymbol{h}_l^T \cdot \boldsymbol{u})}{\sum_{i=1}^{L} exp(\boldsymbol{h}_i^T \cdot \boldsymbol{u})}. \quad (2)$$

Then, the embedding vector $\boldsymbol{e} \in \mathbb{R}^d$ is generated by the weighted sum between the normalized attention weights $w_l$ and $\boldsymbol{x}_l$ as

$$\boldsymbol{e} = \sum_{l=1}^{L} \boldsymbol{x}_l w_l. \quad (3)$$

Hence, an utterance-level feature focused on each frame is encoded. Additionally, based on this process, the MHAP is conducted by performing several linear projections on each input [25].

### 2.3. Previous cross attention and masking methods

Our proposed cross-self-attention and masking methods are inspired by the studies conducted by Lee et al. [18] and Bao et al. [19], respectively.

In image-text matching, cross attention was proposed to identify the appropriate text appearing in an input image [18]. The inputs were encoded in both image-text and text-image formulations. Then, the cross attention was applied to both pairs for obtaining more accurate weights than that obtained with just one attention mechanism.

In person re-identification, masking method and attention mechanisms were applied. These were used to solve the problem of the neglected dissimilarities between the source and the target [19]. In the attention process between the source and the target, a masking matrix of integer [1 or -1], according to the label was used.

## 3. Masked Cross Self-Attentive Encoding

### 3.1. Model architecture

The proposed model builds on previous research on the speaker embedding encoding method based on MLA [14]. The modified ResNet model is trained for speaker classification in an end-to-end training process using several pooling layers.

Table 1: *Proposed model architecture applying MCSAE (D: dimension of input feature, L: length of input feature, N: number of speakers, SE: speaker embedding)*

| Layer | Modified ResNet-34 | Output Size | Embedding Size |
|---|---|---|---|
| conv-1 | $7 \times 7, 32$, stride 1 | $D \times L \times 32$ | - |
| pooling-1 | avg. pooling | - | $1 \times 32\ (\boldsymbol{P}_1)$ |
| res-1 | $\begin{bmatrix} 3 \times 3, 32 \\ 3 \times 3, 32 \end{bmatrix} \times 3$ | $D \times L \times 32$ | - |
| pooling-2 | avg. pooling | - | $1 \times 32\ (\boldsymbol{P}_2)$ |
| mcsae-1 | MCSAE | - | $32 \times 32\ (\boldsymbol{z}_1)$ |
| res-2 | $\begin{bmatrix} 3 \times 3, 64 \\ 3 \times 3, 64 \end{bmatrix} \times 4$ | $D/2 \times L/2 \times 64$ | - |
| pooling-3 | avg. pooling | - | $1 \times 64\ (\boldsymbol{P}_3)$ |
| mcsae-2 | MCSAE | - | $32 \times 64\ (\boldsymbol{z}_2)$ |
| res-3 | $\begin{bmatrix} 3 \times 3, 128 \\ 3 \times 3, 128 \end{bmatrix} \times 6$ | $D/4 \times L/4 \times 128$ | - |
| pooling-4 | avg. pooling | - | $1 \times 128\ (\boldsymbol{P}_4)$ |
| mcsae-3 | MCSAE | - | $64 \times 128\ (\boldsymbol{z}_3)$ |
| res-4 | $\begin{bmatrix} 3 \times 3, 256 \\ 3 \times 3, 256 \end{bmatrix} \times 3$ | $D/8 \times L/8 \times 256$ | - |
| pooling-5 | avg. pooling | - | $1 \times 256\ (\boldsymbol{P}_5)$ |
| mcsae-4 | MCSAE | - | $128 \times 256\ (\boldsymbol{z}_4)$ |
| matmul | - | - | $1 \times 256\ (\boldsymbol{Z})$ |
| concat | - | - | $1 \times 512\ (\boldsymbol{C})$ |
| fc-1 | $512 \times 512$ | - | - |
| fc-2 | $512 \times 512$ | | - |
| fc-3 | $512 \times 512$ | - | $512 \times 1\ (\boldsymbol{SE})$ |
| output | $512 \times N$ | - | - |

The proposed model architecture is modified by using a standard ResNet-34 model [8] and is added MCSAE after each pooling, as shown in Figure 1 and Table 1. The proposed model has 4 residual layers, 16 residual blocks, and half the number of channels of a standard ResNet-34. Each residual block consists of convolution layers, batch normalizations, and leaky ReLU activation functions (LReLU). Especially, the output features of each residual layer is encoded the speaker embedding in order, from low-level representation information to high-level representation information.

The output features ($\boldsymbol{P}_i, \boldsymbol{P}_{i+1}$) of the two previous pooling layers are used as inputs to the $i^{th}$ MCSAE. As shown below

$z_i$, which refers to the $i^{th}$ segment matrix of the attention matrix $Z$ is generated by applying the random masking regularization module and cross self-attention module in the MCSAE:

$$z_i = MCSAE_i(P_i, P_{i+1}) \ (0 \le i \le 4). \quad (4)$$

Here, $z_i$ the output of each MCSAE is used to generate an attention matrix $Z$ of $1 \times 256$ size using matrix product calculation in a *matmul* layer as

$$Z = P_1 \times z_1 \times z_2 \times z_3 \times z_4. \quad (5)$$

To match the dimension, an embedding $P_1$ of $1 \times 32$ size extracted from the *pooling-1* layer is used for the matrix product. Using the $P_1$ matrix allows dimensional matching without increasing the parameters.

In the *concat* layer, embedding $P_5$ of $1 \times 256$ size extracted from the *pooling-5* layer is concatenated with attention matrix $Z$. The embedding $P_5$ is standard embedding in ResNet without the MCSAE. As a result, an embedding $Y$ of $1 \times 512$ size is encoded as

$$C = concat(Z, P_5). \quad (6)$$

Finally, the concatenated embedding $C$ is encoded into fully-connected layers (*fc* layer) and output layer representing the speaker classes (*output* layer). Through this process, a 512-dimensional speaker embedding is extracted.

### 3.2. MCSAE

*3.2.1. Cross self-attention module*

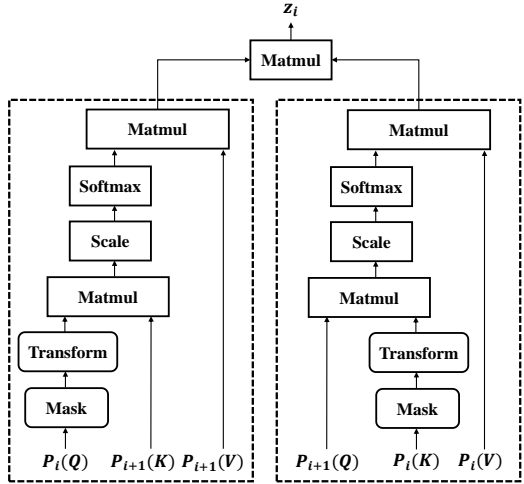

Figure 2: *Overview of the proposed MCSAE (dashed boxes: each self-attention module, matmul: matrix multiplication)*

The MCSAE employs two main proposed modules: 1) a cross self-attention module and, 2) a random masking regularization module. They aim to encode the segment matrix $z_1$ that generates the attention matrix $Z$. The MCSAE is based on the scaled dot-product attention mechanism used in [16]. We assume that the feature $P_i$ is a step preceding feature $P_{i+1}$ and they are closely related to each other, which is further emphasized by the attention mechanism. Therefore, the cross self-attention module is able to train the interdependence between the feature $P_i$ and feature $P_{i+1}$.

As depicted in Figure 2, the MCSAE consists of two input pairs performing cross self-attention. The first self-attention input consists of $P_i$ (query vector, $Q$), $P_{i+1}$ (key vector, $K$), and $P_{i+1}$ (value vector, $V$). After the scaled dot-product operation between $Q$ and $K$, self-attention is performed to the target $V$ as (so, $P_{i+1}$ is the attention target)

$$attention(Q, K, V) = softmax\left(\frac{Q^T K}{\sqrt{d_k}}\right) V^T. \quad (7)$$

Before the scaled dot-product operation, a random masking regularization module is applied to feature $P_i$ as shown in Figure 3. Then, a transform layer is applied to masked $P_i$. Here, an input feature $P_i = [p_1, p_2, ..., p_c, ..., p_C]$ of length $L$ ($p_c \in \mathbb{R}^1$) is fed into the transform layer to obtain $H = [h_1, h_2, ..., h_c, ..., h_C]$ ($h_c \in \mathbb{R}^1$) using LReLU activation function with slope $\lambda$ as

$$h_c = max\{\lambda(Wp_c + b), (Wp_c + b)\}. \quad (8)$$

Next, scaling to the value of $\sqrt{d_k}$ ($d_k$ is the dimension of $K$) is performed and normalization is applied using the softmax function. The computed matrix is multiplied by $V$ and self-attention is finally conducted.

Conversely, the second self-attention input consists of $P_{i+1}$ ($Q$), $P_i$ ($K$), and $P_i$ ($V$). As $P_i$ is the attention target, the scaled dot attention mechanism is performed the same as earlier. The matrix $z_i$ is encoded using matrix multiplication for the output of the masked cross self-attention as

$$z_i = attention_{1st}(P_i, P_{i+1}, P_{i+1}) \times \\ attention_{2nd}(P_{i+1}, P_i, P_i)^T. \quad (9)$$

*3.2.2. Random masking regularization module*

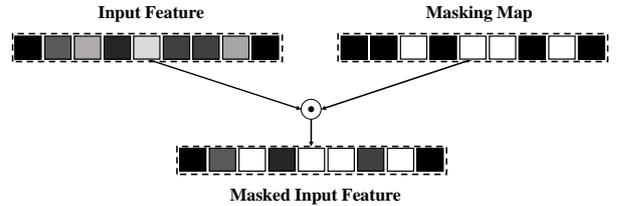

Figure 3: *Overview of the proposed random masking regularization module (The whiter the color, the closer the value is to zero)*

A random masking regularization module prevents overfitting in attention process of the MCSAE layer, as depicted in Figure 3. The masking map consists of random integers, [0 or 1], according to the value of the adaptive scaling factor, which determines the range of masking that is updated by training. As the value increases, the range of the masking widens. Then, masking is performed to input the feature and element-wise multiplication. The masked value was filled with zero.

## 4. Experiments

### 4.1. Dataset setup

In this study, we trained the proposed model using the VoxCeleb2 dataset [27], which contained over 1 million utterances from 5,994 celebrities. These are large-scale text-independent SV datasets collected from YouTube. We evaluated all the experiments using the VoxCeleb1 evaluation dataset containing 40 speakers and 37,220 pairs of official test protocol [28].

### 4.2. Experimental setup

The input feature vectors was 64-dimensional log Mel-filterbank energies of width 25 ms and shift size 10 ms, which were mean-variance normalized over a sliding window of up to 3 s. For each training step, 12 s interval was extracted from each utterance using cropping or padding. In training, we also used the SpecAugment method to perform time and frequency masking on input features [29].

For parameters training, we used the standard stochastic gradient descent optimizer with a momentum of 0.9, a softmax loss function, and a weight decay of 0.0001 at an initial learning rate of 0.1, reduced by a 0.1 decay factor on the plateau. A 96 mini-batch size was used, and early stopping in 200 epochs was performed. The initial adaptive scaling factor was 0.5 in the random masking regularization.

Our proposed model was implemented in an end-to-end manner using PyTorch [30]. It was not used additional methods after extracting the speaker embedding such as [10, 14]. From the trained model, we extracted a speaker embedding and evaluated it using cosine similarity metrics: equal error rate (EER, %) performance and minimum detection cost function (minDCF).

### 4.3. Experimental results

We experimented with the proposed model using three types of comparisons. The first is a comparison with previous self-attentive encoding in Table 2. The second compares performance according to the dimension of speaker embedding in Table 3. The third is a comparison with various state-of-the-art encoding methods in Table 4.

Table 2: *Experimental results compared with previous encoding methods including SAP (Dim: dimension of speaker embedding).*

| Model | Encoding method | Dim | EER | minDCF |
|---|---|---|---|---|
| ResNet-34 | GAP | 256 | 4.57 | 0.2659 |
| | SAP | 256 | 4.24 | 0.2642 |
| | MLA-SAP | 512 | 3.49 | 0.1915 |
| | MCSAE | 512 | 2.63 | 0.1453 |

Table 2 shows the results according to the modifications of ResNet-34 up to the proposed MCSAE method. We applied GAP and SAP methods to ResNet-34. In this case, 256-dimensional speaker embedding was extracted in the last residual layer. Based on MLA, the SAP was performed on the output features of each residual layer (MLA-SAP). Next, the proposed MCSAE method was tested. The results showed that the proposed MCSAE method performed better than the previous self-attentive encoding methods.

Table 3 shows the results according to the dimension of speaker embedding, using the same model with the MCSAE encoding method. The results showed best performance when the dimension was deep such as 512-dimensional speaker embedding.

Table 3: *Experimental results according to dimension of speaker embedding*

| Model | Encoding method | Dim | EER | minDCF |
|---|---|---|---|---|
| ResNet-34 | MCSAE | 64 | 2.65 | 0.1456 |
| | MCSAE | 128 | 2.71 | 0.1481 |
| | MCSAE | 256 | 2.66 | 0.1488 |
| | MCSAE | 512 | 2.63 | 0.1453 |

In contrast, Table 4 shows the results of the comparison with the state-of-the-art encoding methods. Here, we focused on speaker embedding encoding methods using a CNN-based model with the softmax loss function. These models were proposed for using various approaches such as TAP [27], NetVLAD [11], and GhostVLAD [11]. In addition, SAP-derived encoding methods were compared such as MHAP [25], implicit phonetic attention (IPA) [26], including SAP [9]. The results showed that the proposed MCSAE method was comparable to various state-of-the-art encoding methods.

Table 4: *Experimental results compared with state-of-the-arts encoding methods*

| Model | Encoding method | Dim | EER |
|---|---|---|---|
| ResNet-34 [9] * | SAP | 128 | 5.51 |
| VGG [25] * | MHAP | 512 | 4.00 |
| ResNet-34 [27] | TAP | 512 | 5.04 |
| ResNet-50 [27] | TAP | 512 | 4.19 |
| Thin-ResNet-34 [11] | NetVLAD | 512 | 3.57 |
| Thin-ResNet-34 [11] | GhostVLAD | 512 | 3.22 |
| ResNet-17 [26] | PBN-AP | 128 | 2.23 |
| ResNet-34 (**ours**) | MCSAE | 512 | **2.63** |

\* These models used VoxCeleb1 training dataset, which is smaller than the VoxCeleb2 dataset

## 5. Conclusions

In this study, we proposed a new SAP-derived method for speaker embedding encoding called MCSAE. The model was focused on both high-level and low-level layers in the ResNet architecture, in order to encode a more informative speaker embedding. In the MCSAE, the cross self-attention module improved the concentration of the speaker information by training the interdependence among the features of each residual layer. A random masking regularization module prevented overfitting in the attention process of the MCSAE. The experimental results using the VoxCeleb1 evaluation dataset showed that the proposed MCSAE improved performance when compared with previous self-attentive encoding and state-of-the-art encoding methods.

## 6. Acknowledgements

This work was supported by Institute for Information & communications Technology Promotion (IITP) grant funded by the Korea government (MSIT) (No.2017-0-01772, Development of QA systems for Video Story Understanding to pass the Video Turing Test)